\documentclass[aps,prd,12pt,notitlepage,nofootinbib,tightenlines]{revtex4-1}
\usepackage{amsmath}
\usepackage{bm}
\usepackage{times}
\usepackage{braket}
\usepackage{color}
\usepackage{epsfig}
\usepackage{slashed}
\usepackage{hyperref}
\newcommand{\beq}{\begin{eqnarray}}
\newcommand{\eeq}{\end{eqnarray}}
\newcommand{\non}{\nonumber\\ }

\newcommand{\etac}{\eta_c }
\newcommand{\etar}{\eta^\prime }
\newcommand{\etap}{\eta^{(\prime)} }

\newcommand{\jpsi}{J/\Psi}

\newcommand{\psl}{ p \hspace{-2.4truemm}/ }
\newcommand{\nsl}{ n \hspace{-2.2truemm}/ }
\newcommand{\vsl}{ v \hspace{-2.2truemm}/ }

\newcommand{\cala}{ {\cal A} }
\newcommand{\calb}{ {\cal B} }

\def \cpc{ Chin. Phys. C  }

\def \csb{ Chin. Sci. Bull. }

\def \epjc{ Eur. Phys. J. C }
\def \jpg{  J. Phys. G }
\def \npb{  Nucl. Phys. B }
\def \plb{  Phys. Lett. B }
\def \ppnp{ Prog.Part. $\&$ Nucl. Phys. }
\def \prd{  Phys. Rev. D }
\def \prl{  Phys. Rev. Lett.  }

\def \zpc{  Z. Phys. C }
\def \jhep{ JHEP }
\def \rmp{ Rev. Mod. Phys. }
\definecolor{Red}{rgb}{1.,0.,0.}

\definecolor{Blue}{rgb}{0.,0.,1.}

\definecolor{nicered}{rgb}{0.7,0.1,0.1}
\definecolor{nicegreen}{rgb}{0.1,0.5,0.1}
\newcommand{\orcid}[1]{\thanks{\href{http://orcid.org/#1}{ORCID: #1}}}

\hypersetup{colorlinks,citecolor=nicegreen,linkcolor=nicered}
\begin{document}

\title{\boldmath $B_{(s)} \to \eta_c(P,V) $ decays and effects of the next-to-leading order contributions in the perturbative QCD approach}
\author{Zhen-Jun Xiao$^{1,2}$ } \email{xiaozhenjun@njnu.edu.cn}\orcid{0000-0002-4879-209X}
\author{Da-Cheng Yan$^{3}$}   \email{1019453259@qq.com}
\author{Xin Liu$^{4}$}   \email{liuxin@jsnu.edu.cn}\orcid{0000-0001-9419-7462}
\affiliation{$^1$ Department of Physics and Institute of Theoretical Physics,
Nanjing Normal University, Nanjing, Jiangsu 210023, China}
\affiliation{$^2$ Jiangsu Key Laboratory for Numerical Simulation of Large Scale Complex
Systems, Nanjing Normal University, Nanjing, Jiangsu 210023, China}
\affiliation{$^3$ School of Mathematics and Physics, Changzhou University, Changzhou 213164, China}
\affiliation{$^4$ Department of Physics, Jiangsu Normal University, Xuzhou 221116, China}
\date{\today}
\begin{abstract}
By employing the perturbative QCD (PQCD) factorization approach, we studied the sixteen $B/B_s \to \eta_c (\pi, K, \etap,\rho,K^*,\omega,\phi)$ decays with the inclusion of the currently known
next-to-leading order (NLO) contributions. We found the following main points:
(a) for the five measured $B \to \eta_c (K,K^*)$ and $B_s\to \eta_c\phi$ decays, the NLO contributions can provide $ (80-180)\%$ enhancements to the
leading order (LO) PQCD predictions of their branching ratios, which play an important role to help us to
interpret the data;
(b) for the seven  ratios $R_{1,\cdots,7}$ of the branching ratios defined among the properly selected pair of the considered decay modes,
the PQCD predictions for the values of $R_{3,4,5}$ agree well with those currently available  measurements from BaBar and Belle Collaboration;
(c) for $B^0 \to \eta_c  K_S^0$ decay, the PQCD predictions for both the direct and mixing induced CP asymmetries do agree very well
with the measured values within errors; and
(d) the PQCD predictions for ratios $R_{1,2}$ and $R_{6,7}$ also agree with the general expectations and will be tested by the future
experiments.
\end{abstract}

\pacs{13.25.Hw, 12.38.Bx, 14.40.Nd}

\vspace{1cm}

\maketitle
{\bf \rm Key Words:}{$B$ meson decays; The PQCD factorization approach;
Branching ratios; CP asymmetries}

\section{Introduction}

Analogous to the well-studied    $B \to \jpsi  (P,V) $  decays,  the $B_{(s)} \to \eta_c (P,V) $  decays also play an important role in our efforts
to measure the Cabibbo-Kobayashi-Maskawa (CKM)  matrix element $V_{cb}$  and  the CP violating phase  $\phi_s$,
and have drawn great attention for many years.
Very recently,   the decay $B^0_s \to \eta_c \phi$ was measured by LHCb collaboration~\cite{lhcb17}:
\begin{eqnarray}
{\calb } (B^0_s \to \eta_c \phi)=(5.01 \pm 0.53 \pm 0.27 \pm 0.63) \times 10^{-4}.
\end{eqnarray}
For other similar decays, such as the four $B \to \eta_c K^{(*)}$ decay modes, their branching ratios have also been measured by Belle~\cite{belle18,belle03}
and BaBar~\cite{Babar04,Babar05,Babar07,Babar08} Collaborations. Furthermore, the direct and mixing induced CP violating
asymmetries of the decay $B^0 \to \eta_c K_{S}^0$ are also given in PDG~\cite{Aubert:2009aw,pdg2018} :
\begin{eqnarray}
A^{\rm dir}_{\rm CP}( B^0 \to \eta_c K_S^0)&=&0.08 \pm 0.13,  \non
A^{\rm mix}_{\rm CP}( B^0 \to \eta_c K_S^0)&=&0.93 \pm 0.17.
\label{eq:acpe1}
\end{eqnarray}

\begin{figure}[bh]
\vspace{-3cm}
\centerline{\epsfxsize=19cm \epsffile{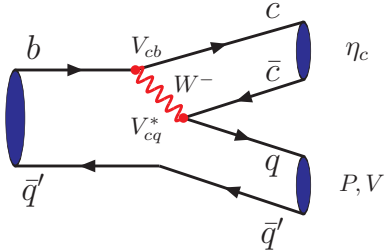}}
\vspace{-20cm}
\caption{ (Color online) The leading  Feynman diagrams for  $B \to \eta_c (P,V)$ decays, where $P=(\pi,K,\eta,\etar)$ and $V=(\rho,K^*,\phi,\omega)$
are light charmless mesons. }
\label{fig:fig1}
\end{figure}

On the theory side, such kinds of $B/B_s$  meson decays have been studied intensively
by employing rather different theocratical methods, such as the naive factorization approach(NFA)~\cite{nfa2},
the QCD-improved factorization(QCDF) approach~\cite{chay99,cheng01,QCDF03,QCDF041,QCDF042},
the final-state interactions (FSI) \cite{fsi1,fsi2,fsi3}, and the light-cone sum rules(LCSR)~\cite{lcsr1,lcsr2,lcsr3}.
At the quark level, all considered decay modes are induced by the $b \to c\bar{c} q (q=d,s)$ transitions in the framework of  the standard model(SM)
and belong to the color-suppressed category,  as illustrated by the leading Feynman diagram in Fig.~\ref{fig:fig1}.
In  Ref.~\cite{QCDF03}, for example,  the authors studied $B\to \eta_c K^0$ decay and found a small decay rate:
\beq
\calb(B \to \eta_c K^0) = (1.4-1.9) \times 10^{-4},
\eeq
which is  only abour $20\%$ of the world average $(8.0\pm 1.2)\times 10^{-4}$ as given in PDG 2018 \cite{pdg2018}.
In Ref.~\cite{fsi1},  on the other hand, the author studied $B^0\to \eta_c K^{*}$ decay and found that  the FSI correction could be
comparable with the contribution  from the naive factorizable amplitude,  and the prediction  with  the inclusion of  the FSI part  was increased significantly
to the value
\beq
\calb(B \to \eta_c K^*) = (4.83-6.94) \times 10^{-4},
\eeq
which is  well consistent with the experimental data.

In the PQCD approach ~\cite{li2003,nlo05,pqcd1,pqcd2},  fortunately,  the hard spectator amplitudes can be calculated reliably.
By employing the PQCD approach,  many  $B_{(s)} \to  (c\bar{c}) M$ decays have been
studied at the leading order (LO) or the partial next-to-leading order (NLO),
such as $B \to (J/\psi, \eta_c)K^{(*)}$ decays ~\cite{pqcd05,cpc10,liu14} and even
the excited states $B \to \psi(2S) V$~\cite{zhou17} and $B \to (\psi(2S), \eta_c(2S))(\pi,K)$~\cite{zhang17}.
Most theocratical predictions as presented in Refs.~\cite{pqcd05,cpc10,liu14,zhou17,zhang17}  are  well consistent  with currently available
experimental measurements.

In this paper,  we will make a comprehensive study for the sixteen $B_{(s)} \to \eta_c (P,V)$ ( where
$P=(\pi,K,\eta^{(\prime)})$ and $V=(\rho,K^*,\omega,\phi)$ are the light charmless mesons ) decays by employing the PQCD approach.
Apart from the full LO contributions, the  NLO vertex corrections
are also taken into account. Besides, the  NLO twist-2 and twist-3 contributions to
the form factors of $B_{(s)} \to P$ transitions are also included in $B_{(s)} \to \eta_c  P$ decays.

This paper is organized as follows. In Sec.~\ref{sec:lo-nlo}, we give a brief review about the PQCD
factorization approach and  then calculate analytically the relevant Feynman diagrams and present the
various decay amplitudes for the considered decay modes at the LO and NLO level.
In Sec.~\ref{sec:n-d}, we will  show the PQCD predictions for the branching ratios and CP violating
asymmetries of all sixteen $B_{(s)} \to \eta_c (P,V)$ decays and  make some phenomenological discussions about these results.
A short summary is given in the last section.

\section{ Decay amplitudes at LO and NLO level}\label{sec:lo-nlo}

In the PQCD approach, we treat the $B$ meson\footnote{In this paper, the term $B$ meson generally denotes  the $B_{u,d}$ meson or the $B_s$ meson.}
as a heavy-light system and consider it at rest for simplicity.
By employing the light-cone coordinates, we define the $B$ meson with momentum
$p_1$, the emitted meson $M_2=\eta_c$  with the momentum $p_2$
along the direction of $n=(1,0,{\bf 0}_{\rm T})$, and the
recoiled meson $M_3=(P,V)$  with the momentum $p_3$ in the direction of $v=(0,1,{\bf 0}_{\rm T})$
 ( here $n$ and $v$ are the light-like dimensionless unit vectors) ,  in  the following form:
\beq
p_1 = \frac{m_B}{\sqrt{2}} (1,1,{\bf 0}_{\rm T}), \quad
p_2 = \frac{m_B}{\sqrt{2}} (1-r_3^2,r_2^2,{\bf 0}_{\rm T}), \quad
p_3 = \frac{m_B}{\sqrt{2}} (r_3^2,1-r_2^2,{\bf 0}_{\rm T}).
\eeq
The longitudinal polarization vector of the final state vector meson can then be parameterized as:
\begin{eqnarray}
\epsilon_3^L=\frac{1}{\sqrt{2(1-r_2^2)}r_3}(-r_3^2,1-r_2^2,{\bf 0}_{\rm T}),
\end{eqnarray}
where $r_2=m_{\eta_c}/m_{B}$ and  $r_3=m_3/{m}_{B}$ are the ratios of the meson masses,
${m}_B$ is the initial $B$ meson mass,   $m_{\eta_c}$ and $m_3$ are the masses  of the final state mesons.
The momenta $k_i(i=1,2,3)$ carried by the light anti-quark in the initial $B$ and the final $M_{2,3}$ mesons are chosen as follows:
\beq
k_1 &=& \left (x_1   \frac{{m}_B}{\sqrt{2}}, 0, {\bf k}_{\rm 1T} \right ), \non
k_2 &=& \left ( x_2(1-r_3^2)\frac{{m}_B}{\sqrt{2}} , x_2r_2^2 \frac{{m}_B}{\sqrt{2}}, {\bf k}_{\rm 2T} \right ), \non
k_3 &=& \left  ( x_3r_3^2 \frac{{m}_B}{\sqrt{2}}, x_3(1-r_2^2) \frac{{m}_B}{\sqrt{2}}, {\bf k}_{\rm 3T} \right ),
\eeq
where $x_i$ with $i=(1,2,3)$ are the momentum fraction of the light anti-quark in $B$ and final state $M_{2,3}$.

 Based on above definitions, the decay amplitude for  the considered $B \to  \eta_c M_3$ decays
can be written symbolically in the following form,
\beq
\cala(B \to \eta_c M_3) &\sim & \int\!\! d x_1 d x_2 d x_3 b_1 d b_1 b_2 d b_2 b_3 d b_3  \non
&& \cdot \mathrm{Tr}\left [ C(t) \Phi_{B}(x_1,b_1) \Phi_{\eta_c}(x_2,b_2) \Phi_{M_3}(x_3, b_3) H(x_i, b_i, t) S_t(x_i)\, e^{-S(t)} \right ], \quad \label{eq:a2}
\eeq
in which, $b_i$ is the conjugate space coordinate of transverse momentum $k_{\rm iT}$,
$C(t)$ stands for the Wilson coefficients evaluated at the scale $t$,  and $\Phi_i$ denotes the
wave functions of the initial and final state mesons. The kernel $H(x_i,b_i,t)$ describes the hard dynamics
associated with the effective ``six-quark interaction" with a hard gluon.
The Sudakov factors $e^{-S(t)}$ and $S_t(x_i)$ together can suppress the soft dynamics in the endpoint region
effectively~\cite{li2003}.

\subsection{ Wave functions and decay amplitudes}\label{sec:wf}

For the wave function of the $B$ meson,  we adopt its wave function
as being widely used,  for example, in Refs.~\cite{li2003,ali07,vv15}
\beq
\Phi_{B}&=& \frac{1}{\sqrt{6}} (\psl_{B} +{m}_{B}) \gamma_5 \phi_{B} ({\bf k}),
\label{eq:bsmeson}
\eeq
where  the distribution amplitude (DA) $\phi_{B}$ can be parameterized  in the following form\footnote{Very recently, a new method was proposed to calculate the $B$-meson
light-cone distribution amplitude from Lattice QCD. The interested reader could refer to Ref.~\cite{Wang:2019msf} for detail.} \cite{li2003}:
\beq
\phi_{B}(x,b)&=& N_{B} x^2(1-x)^2 \exp \left  [ -\frac{m_{B}^2\ x^2}{2 \omega_{B}^2}
-\frac{1}{2} (\omega_{B} b)^2\right],
\label{phib}
\eeq
with $\omega_{B}$ being the shape parameter.  According to the discussions in Ref.~\cite{li2003,ali07,jpg06},
we here take $\omega_{B} =0.40 \pm 0.04$ GeV for the $B_{u,d}$ mesons \cite{li2003},   and  $\omega_{B} =0.50 \pm 0.05$ GeV for the $B_s$ meson \cite{ali07,jpg06}.
The normalization factor $N_{B}$ will be determined through the normalization condition: $\int \phi_{B}(x,b=0)  d x =f_{B}/(2\sqrt{6})$.

For the pseudoscalar charmonium state $\eta_c$,  its  wave function can be written in the form of
\beq
\Phi_{\eta_c}(x)&=& \frac{i}{\sqrt{2 N_c}}\gamma_5 \bigg\{\psl \phi_{\eta_c}^v(x)+m_{\eta_c} \phi_{\eta_c}^s(x)\bigg\}\;,
\eeq
where the twist-2 and twist-3 asymptotic distribution amplitudes (DAs), $\phi^v$ and $\phi^s$, can be read as~\cite{etacda},
\beq
\phi_{\eta_c}^v(x)&=&9.58\frac{f_{\eta_{c}}}{2\sqrt{2N_c}}x(1-x) \left[\frac{x(1-x)}{1-2.8x(1-x)}\right]^{0.7}\;, \non
\phi_{\eta_c}^s(x)&=&1.97\frac{f_{\eta_{c}}}{2\sqrt{2N_c}} \left[\frac{x(1-x)}{1-2.8x(1-x)}\right]^{0.7}\;.  \label{eda}
 \eeq

For the light pseudo-scalar  states
$M=(\pi, K,\eta_q,\eta_s)$, their wave functions are the same
ones as those in Refs.~\cite{bf90,pball98,pball99,bl04,kmm04,bz05,liwf,wuwf,fan2013}:
\beq
\Phi_{M}(x)\equiv \frac{1}{\sqrt{6}}\gamma_5 \left [ \psl \phi^{A}_{M}(x)
+m_{0}^{{M}} \phi_{M}^{P}(x)+ \zeta m_0^{{M}} (\nsl \vsl -1)\phi_{M}^{T}(x)\right ], \label{eq:phip}
\eeq
where $m_0^{M}$ is the chiral mass of the relevant state
$M$,  $p$ and $x$ are the momentum and
the fraction of the momentum of $M$. The parameter $\zeta=1$ or $-1$ when the
momentum fraction of the quark (anti-quark) of the meson is set to be $x$.
The DAs of the state $M$ can be found easily, for example,  in Refs.~\cite{pball98,pball99,fan2013,liwf,wuwf}:
\beq
\phi_{M}^A(x) &=&  \frac{3 f_M}{\sqrt{6} } x (1-x)
    \left[1+a_1^{M}C^{3/2}_1(t)+a^{M}_2C^{3/2}_2(t)+ a_4^M C_4^{3/2}(t) \right],\label{eq:piw1}\\
\phi_M^P(x) &=&   \frac{f_M}{2\sqrt{6} }
   \left \{ 1+\left (30\eta_3-\frac{5}{2}\rho^2_{M} \right ) C^{1/2}_2(t)
   -3\left [ \eta_3 \omega_3 + \frac{9}{20}\rho_M^2\left ( 1 + 6a_2^M\right)C_4^{1/2}(t)\right]
   \right \}, \ \
\label{eq:piw2}   \\
\phi_M^T(x) &=&  \frac{f_M(1-2x)}{2\sqrt{6} }
   \left\{ 1+6\left [ 5\eta_3-\frac{1}{2}\eta_3\omega_3-\frac{7}{20}\rho^2_M
   -\frac{3}{5}\rho^2_M a_2^{M} \right ]
   \left (1-10x+10x^2\right )\right \},\quad
   \label{eq:piw3}
\eeq
where $t=2x-1$, $f_M$ and $\rho_M$ are the decay constant and  the mass ratio with the definition of
$\rho_M=(m_\pi/m_0^\pi,m_K/m_0^K$, $m_{qq}/m_0^{\eta_q},m_{ss}/m_0^{\eta_s})$.
The parameters $(m_{qq},m_0^{\eta_q},m_{ss},m_0^{\eta_s})$  have been defined in Eq.~(23) of Ref.~\cite{ckl06}.
The explicit expressions of those Gegenbauer polynomials $C_1^{3/2}(t)$ and $C_{2,4}^{1/2,3/2}(t)$ in Eqs.~(\ref{eq:piw1},\ref{eq:piw2})
can be found for example in Eq.~(20) of Ref.~\cite{xiao08b}.
The Gegenbauer moments $a_i^M$ and other input parameters are the same as those in Refs.~\cite{pball99,bl04,kmm04,bz05}
\beq
a^{\pi,\eta_q,\eta_s}_1 &=& 0,\quad a^K_1  = 0.06, \quad a^{\pi,K}_2=  a^{\eta_q,\eta_s}_2= 0.25\pm 0.15,  \non
a^{\pi, K,\eta_q,\eta_s}_4  &=& -0.015, \quad \eta_3= 0.015, \quad \omega_3=-3.0.
\label{eq:aim}
\eeq

For $\eta-\etar$ mixing, we adopt  the quark-flavor basis:
$\eta_q= (u\bar u +d\bar d)/\sqrt{2}$ and $\eta_s=s\bar{s}$ as being used for example in Refs.~\cite{nlo05,ckl06,fan2013}.
The physical $\eta$ and $\etar$ can then be written in the form of
\beq
\left(\begin{array}{c} \eta \\ \eta^{\prime}\end{array} \right)= \left ( \begin{array}{cc}
\cos\phi & -\sin\phi\\ \sin\phi & \cos\phi\\ \end{array} \right)
\left(\begin{array}{c} \eta_q \\ \eta_s\end{array} \right),\label{eq:e-ep2}
\eeq
where the angle $\phi$  is the mixing angle between $\eta_q$ and $\eta_s$.
The relation between the decay constants $(f_\eta^q, f_\eta^s,f_{\etar}^q,f_{\etar}^s)$ and $(f_q,f_s)$ can be found for example
in Ref.~\cite{fan2013}. The chiral masses $m_0^{\eta_q}$ and $m_0^{\eta_s}$ have been defined
in Ref.~\cite{ckl06} by assuming the exact isospin symmetry $m_q=m_u=m_d$.
The parameters $(f_q, f_s)$ and mixing angle $\phi$ in Eq.~(\ref{eq:e-ep2}) have been extracted from the data \cite{fks98,fks99}
\beq
f_q=(1.07\pm 0.02)f_{\pi},\quad f_s=(1.34\pm 0.06)f_{\pi},\quad \phi=39.3^\circ \pm 1.0^\circ.
\eeq
With $f_\pi=0.13$ GeV,  the chiral masses $m_0^{\eta_q}$ and $m_0^{\eta_s}$ will take the values of
$m_0^{\eta_q}=1.07$ GeV and $m_0^{\eta_s}=1.92$ GeV \cite{ckl06}.
Analogously,  we adopt the ideal form  for the $\omega-\phi$ mixing as $\omega = (u\bar u +  d\bar d)/\sqrt{2}$ and $\phi= s \bar s$.

For the considered $B  \to \eta_c V$ decays, only the longitudinal polarization component
of the involved vector mesons contributes to the decay amplitudes.
Therefore we choose the wave functions of the vector mesons as in Refs.~\cite{ali07,npbvda,bz05b,pball07}:
\begin{eqnarray}
\Phi_{V}^{\|}({p},\epsilon_{L})\,=\,\frac{1}{\sqrt{6}}\left[m_{V}\makebox[0pt][l]{/}
\epsilon_{L}\phi_{V}(x)\,+\,\makebox[0pt][l]{/}\epsilon_{L}\makebox[-1.5pt][l]
{/}{p}\phi_{V}^{t}(x)+m_{V}\phi_{V}^{s}(x)\right],
\label{eq:phiv}
\end{eqnarray}
where $p$ and $m_V$ are the momentum and the mass of the light vector mesons $(\rho,K^*,\phi,\omega)$,
and  $ \epsilon_L $ is the longitudinal polarization vector of these vector mesons.
The twist-2  DA $\phi_V(x)$  and the twist-3 DAs   $\phi^t_V(x)$ and $\phi^s_V(x)$ in Eq.~(\ref{eq:phiv})
can be written in the following form~\cite{npbvda,bz05b,pball07}
\begin{eqnarray}
\phi_{V}(x)  &=& \frac{3f_{V}}{\sqrt{6}}x(1-x)\left[1+a^{\|}_{1V}C_{1}^{3/2}(t) +a_{2V}^{\|}C_2^{3/2}(t)\right],  \label{eq:phivx} \\
\phi^t_V(x) &=& \frac{3f^T_V}{2\sqrt 6}(2x-1)^2, \non
\phi^s_V(x) &=& \frac{3f_V^T}{2\sqrt 6} (1-2x)~,\label{eq:phivxs}
\end{eqnarray}
where $t=2x-1$, $f_{V}$ ($f_{V}^T$  )  is the decay constant of the vector meson with longitudinal  (transverse) polarization.
The Gegenbauer moments in Eq.~(\ref{eq:phivx})  are  the same as those in Refs.~\cite{npbvda,bz05b}:
\begin{eqnarray}
a_{1\rho}^{\|}&=& a_{1\omega}^{\|}=a_{1\phi}^{\|}=0, \quad  a_{1K^*}^{\|}=0.03\pm0.02,\quad  a_{2\phi}^{\|}=0.18\pm 0.08,\non
a_{2\rho}^{\|} &=&a_{2\omega}^{\|}=0.15\pm 0.07,\quad  a_{2K^*}^{\|}=0.11\pm0.09. \label{eq:Geb}
\end{eqnarray}

\subsection{ Example of the LO decay amplitudes}\label{sec:lo-aml}

In the SM,  for the considered $B \to \eta_c (P,V)$ decays induced by
the $b \to q$ transition with $q=(d,s)$,
the weak effective Hamiltonian $H_{eff}$ can be written as\cite{buras96},
\beq
\label{eq:heff}
H_{eff} &=& \frac{G_{F}}{\sqrt{2}}     \Bigg\{ V_{cb} V_{cq}^{\ast} \Big[
 C_{1}({\mu}) O^{c}_{1}({\mu})  +  C_{2}({\mu}) O^{c}_{2}({\mu})\Big]
  -V_{tb} V_{tq}^{\ast} \Big[{\sum\limits_{i=3}^{10}} C_{i}({\mu}) O_{i}({\mu})
  \Big ] \Bigg\} + \mbox{h.c.}
\eeq
where the Fermi constant $G_{F}=1.166 39\times 10^{-5}$ GeV$^{-2}$, and
$V_{ij}$ is the 
CKM matrix element,
$C_i(\mu)$ are the Wilson coefficients and $O_i(\mu)$
are the local four-quark operators. For convenience, the combinations $a_i$ of the
Wilson coefficients are defined as usual~\cite{ali07,vv15}:
\begin{eqnarray}
\label{eq:ai}
&&a_{1}=C_{2}+C_{1}/3,\;\;\;\;\;\;a_{2}=C_{1}+C_{2}/3,\nonumber\\
&&a_{i}=C_{i}+C_{i\pm 1}/3,\,(i=3 - 10) \;.
\end{eqnarray}
where the upper(lower) sign applies, when $i$ is odd(even).

\begin{figure}[tb]
\vspace{-2cm}
\centerline{\epsfxsize=17cm \epsffile{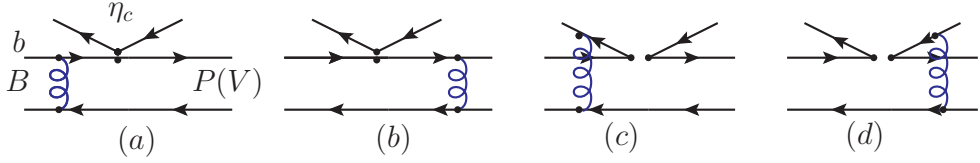}}
\vspace{-20cm}
\caption{ (Color online) The typical Feynman diagrams of $B \to \eta_c (P,V)$ decays in the leading order PQCD approach.  }
\label{fig:fig2}
\end{figure}

In the leading order PQCD approach, as illustrated in Fig.~\ref{fig:fig2}, there are only four types of the Feynman diagrams
contributing to the $B \to \eta_c (P,V)$ decays, which can be classified into two types:
(a) the factorizable emission diagrams ( Fig.~\ref{fig:fig2}(a) and \ref{fig:fig2}(b) ),  and (b)
the nonfactorizable emission  (hard-spectator) diagrams (Fig.~\ref{fig:fig2}(c) and \ref{fig:fig2}(d) ).
By evaluating and combining the contributions from the different Feynman diagrams as illustrated in Fig.~\ref{fig:fig2}, one can get the total
decay amplitudes for the $B \to \eta_c(P,V)$ decays:
\beq
  A(B   \to \eta_c(P,V))& =& V_{cb}V_{cd(s)}^{*}[a_2 {{\cal F}}_{eP(V)}^{LL}+
  C_2 {{\cal M}}_{eP(V)}^{LL}]-V_{tb}V_{td(s)}^{*}\Bigl [(a_3+a_9){\cal F}_{eP(V)}^{LL} \nonumber \\
  &&+(a_5+a_7) {\cal F}_{eP(V)}^{LR}+(C_4+C_{10}) {\cal M}_{eP(V)}^{LL}+(C_6+C_8) {\cal M}_{eP(V)}^{SP}\Bigr]
\label{eq:aa10}
\eeq
where the terms ${\cal F}$ and ${\cal M}$  describes the contributions from the factorizable and nonfactorizable diagrams respectively.
The superscript $LL, LR$ and $SP$ refers to the contributions from the $(V-A)\otimes (V-A)$,
$(V-A)\otimes (V+A)$ and $(S-P)\otimes (S+P)$ operators, respectively.
The explicit expressions of ${\cal F}^{LL,LR}_{eP}, {\cal F}^{LL,LR}_{eV},{\cal M}^{LL,SP}_{eP}$ and
${\cal M}^{LL,SP}_{eV}$ are of the form:
\begin{eqnarray}
{\cal F}_{e P}^{LL}&=&-{\cal F}_{e P}^{LR}=8 \pi C_F {m}^4_{B} {f_{\eta_c}} \int^1_0 dx_1dx_3 \int_0^{\infty}b_1db_1b_3db_3\phi_B(x_1,b_1) \non
 &&\times \Big \{ \Big [ \left [ (1-r_2^2)(1+x_3)-x_3r_2^2 \right ] \phi_{M}^A(x_3)+r_{0}^{M}(1-2x_3)  [\phi_{{M}}^P(x_3)+\phi_{{M}}^T(x_3)]\non
 &&+r_{0}^{M}r^2_2 \left [ (1+2x_3)\phi_{M}^P(x_3)-(1-2x_3)\phi_{M}^T(x_3) \right ] \Big ] \non
 &&\cdot\alpha_s(t_a)h_e(x_1,x_3,b_1,b_3)S_{ab}(t_a)S_t(x_1)\non
 &&+2r_{0}^{M}(1-r_2^2)\phi_{M}^P(x_3)\cdot\alpha_s(t_b)h_e(x_3,x_1,b_3,b_1)S_{ab}(t_b)S_t(x_3) \Big \},
\label{eq:fep1}
\end{eqnarray}
\begin{eqnarray}
{\cal F}_{e V}^{LL}&=&-{\cal F}_{e V}^{LR}=8 \pi C_F m^4_{B} {f_{\eta_c}} \int^1_0 dx_1dx_3
\int_0^{\infty}b_1db_1b_3db_3\phi_B(x_1,b_1) \non
 &&\times \Big \{\sqrt{1-r_2^2} \left [ (r_2^2-1)x_3-1 \right ]\phi_V(x_3)
 -r_V \left [ 1-2x_3+r_2^2(2x_3+1) \right ]\phi^s_V(x_3)\non
 &&-r_V(1-r_2^2)\sqrt{1-r_2^2}(1-2x_3)\phi^t_V(x_3)
 \cdot\alpha_s(t_a)h_e(x_1,x_3,b_1,b_3)S_{ab}(t_a)S_t(x_1)\non
 &&-2r_V(1-r_2^2)\phi_V^s(x_3)\cdot\alpha_s(t_b)h_e(x_3,x_1,b_3,b_1)S_{ab}(t_b)S_t(x_3) \Big \},
\label{eq:fev1}
\end{eqnarray}
\begin{eqnarray}
{\cal M}_{e P}^{LL}&=&{\cal M}_{e P}^{SP}=-\frac{32}{\sqrt{6}} \pi C_F  m^4_{B}\int^1_0 dx_1dx_2dx_3
\int_0^{\infty}b_1db_1b_2db_2\phi_B(x_1,b_1)\non
 &&\times x_3\phi^v_{\eta_c}(x_2,b_2) \Big [ (1-2r_2^2)\phi_{M}^A(x_3)
 -2r_{0}^{M}(1-r_2^2)\phi_{M}^T(x_3) \Big ]\non
 &&\cdot\alpha_s(t_f)h_f(x_1,x_2,x_3,b_1,b_2)S_{cd}(t_f),
\label{eq:mep1}
\end{eqnarray}
\begin{eqnarray}
{\cal M}_{e V}^{LL}&=&-{\cal M}_{e V}^{SP}=\frac{32}{\sqrt{6}} \pi C_F  m^4_{B}\int^1_0 dx_1dx_2dx_3
\int_0^{\infty}b_1db_1b_2db_2\phi_B(x_1,b_1)\non
 &&\times x_3\phi^v_{\eta_c}(x_2,b_2)
 \sqrt{1-r_2^2} \Big [ (1-r_2^2)\phi_V(x_3)-2r_V(1-r_2^2)\phi_V^t(x_3) \Big ]\non
 &&\cdot\alpha_s(t_f)h_f(x_1,x_2,x_3,b_1,b_2)S_{cd}(t_f),
 \label{eq:mev1}
\end{eqnarray}
where $C_F = 4/3$ and $\alpha_s(t_i)$ is the strong coupling constant. In the above functions,
$r_V=m_V/m_B$ and $r_{0}^{M}=m_0^{M}/m_B$ with $m_0^{M}$ the chiral
mass of the pseudoscalar state.
The explicit expression of the Sudakov factors $(S_{ab}(t_a),S_{ab}(t_b),S_{cd}(t_f))$ and $S_t(x_i)$,
 the hard scales $t_i$ with $i=(a,b,f)$ and the hard functions $h_{e,f}(x_i,b_j)$  can be found in Refs.~\cite{zhou17,zhang17,cpc10,pqcd05,etacsf}.

\begin{figure}[tb]
\vspace{-2cm}
\centerline{\epsfxsize=17cm \epsffile{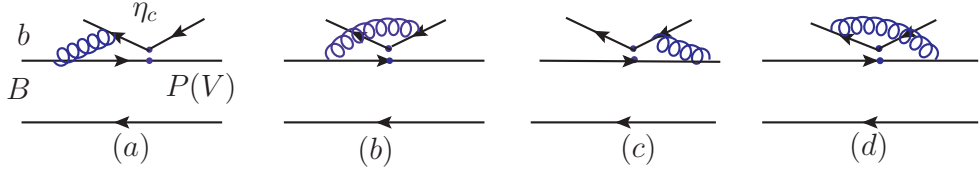}}
\vspace{-20cm}
\caption{The NLO vertex corrections to the factorizable diagrams for the $B \to \eta_c (P,V)$ decays. }
\label{fig:fig3}
\end{figure}

\begin{figure}[tb]
\vspace{-2cm}
\centerline{\epsfxsize=17cm \epsffile{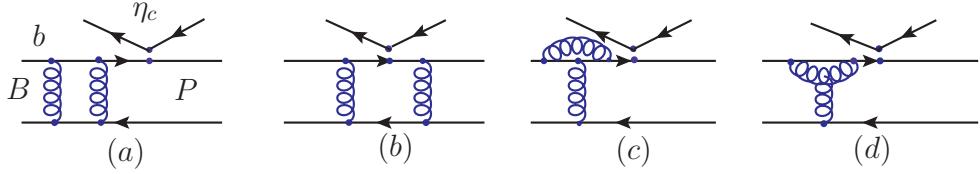}}
\vspace{-20cm}
\caption{ The typical Feynman diagrams of NLO twist-2 and twist-3 contributions to the $B \to P$ transition form factors. }
\label{fig:fig4}
\end{figure}

In this work, beyond the full LO contributions, the following two currently
known NLO corrections to the considered $B \to \eta_c (P,V)$ decays
are also taken into account:
\begin{enumerate}
\item[(1)]
The NLO vertex corrections to the factorizable amplitudes ${\cal F}_{eP}$ and ${\cal F}_{eV}$,  as shown in Fig.~\ref{fig:fig3}.

\item[(2)]
The NLO twist-2 and twist-3 contributions to the form factors of $B \to P$ transitions, as shown in Fig.~\ref{fig:fig4}.

\end{enumerate}

According to Refs.~\cite{QCDF041,pqcd05}, the vertex corrections can be absorbed
into the redefinition of the Wilson coefficients $a_i(\mu)$  by adding an additional term to them:
\beq
a_2 &\rightarrow & a_2 +\frac{\alpha_s}{9\pi}C_2 \left[-18-12\text{ln}(\frac{\mu}{m_b})+f_I\right ],\non
a_3+a_9 &\rightarrow &a_3+a_9+\frac{\alpha_s}{9\pi}(C_4+C_{10})\left [-18-12\text{ln}(\frac{\mu}{m_b})+f_I \right ],\non
a_5+a_7&\rightarrow&a_5+a_7+\frac{\alpha_s}{9\pi}( C_6+C_{8})\left [6+12\text{ln}(\frac{\mu}{m_b})-f_I\right ], \label{eq:vertex}
\eeq
where the function $f_I$ describes the vertex corrections ~\cite{chay99,cheng01}:
\beq
f_I=\frac{2\sqrt{6}}{f_{\etac} } \int \; dx   \phi^v_{\etac}(x) \cdot
\left [  \frac{3(1-2x)}{1-x}\ln[x] -3\pi  i + 3 \ln[1-r_2^2] + \frac{2r_2^2(1-x)}{1-r_2^2 x}\right].
\label{eq:fi1}
\eeq
 For more discussions about the properties of function $f_I$, one can see Refs.~\cite{chay99,cheng01,QCDF03,QCDF041,QCDF042,cpc10}.

The NLO twist-2 and twist-3 contributions to the form factors of $B_{u,d} \to \pi$ transition have been calculated very recently in Refs.~\cite{prd85-074004,cheng14a}.
 We can extend the formulas for $B_{u,d}\to \pi$ transitions as given in Refs.~\cite{prd85-074004,cheng14a}
to the cases for $B_{(s)} \to ( K, \eta_q, \eta_s)$ transition form factors directly,
by  making appropriate replacements  for the corresponding distribution amplitudes and some relevant parameters.
The NLO form factor $f^+(q^2)$ for $B_s \to K$ transition, for example, can be written in the following form:
\beq
f^+(q^2)|_{\rm NLO} &=& 8 \pi m^2_{B_s} C_F \int{dx_1 dx_3} \int{b_1 db_1 b_3 db_3}
\phi_{B_s}(x_1,b_1)\non &&
\hspace{-1.5cm}\times \Biggl \{ r_{0}^K \left [\phi_{K}^{P}(x_3) - \phi_{K}^{T}(x_3) \right ]
\cdot \alpha_s(t_a)\cdot e^{-S_{ab}(t_a)}\cdot S_t(x_3)\cdot h_e(x_1,x_3,b_1,b_3) \non
&&\hspace{-1.5cm}  + \Bigl [ (1 + x_3 \eta)
\left (1 + F^{(1)}_{\rm T2}(x_1,x_3,\mu) \right ) \phi_{K}^A(x_3)
+ 2 r_{0}^K \left (\frac{1}{\eta} - x_3 \right )\phi_{K}^T(x_3)
- 2x_3 r_{0}^K \phi_{K}^P(x_3) \Bigr ] \non
&& \hspace{-1.5cm} \cdot \alpha_s(t_a)\cdot e^{-S_{ab}(t_a)} \cdot S_t(x_3)\cdot h_e(x_1,x_3,b_1,b_3)\non
&& \hspace{-1.5cm} + 2 r_{0}^K \phi_{K}^P(x_3) \left (1 + F^{(1)}_{\rm T3}(x_1,x_3,\mu) \right )
\cdot \alpha_s(t_b)\cdot e^{-S_{ab}(t_b)} \cdot S_t(x_3)\cdot h_e(x_3,x_1,b_3,b_1) \Biggr \},
\label{eq:ffnlop}
\eeq
where $\eta=1-q^2/m_{B_s}^2$ with $q^2=(P_{B_s}-P_3)^2$ and $P_3$ is the momentum of the
meson $M_3$ which absorbed the spectator $\bar{s}$ quark of the $\bar{B}^0_s$ meson, $\mu$ is  the
renormalization scale, the hard scale $t_{a,b}$ are
chosen as the largest scale of the propagators in the hard $b$-quark decay diagrams
\cite{prd85-074004,cheng14a}. The explicit expressions of the threshold Sudakov function
$S_t(x_3)$ and the hard function
$h_e(x_i,b_j)$ can be found in Refs.~\cite{prd85-074004,cheng14a}.
The NLO correction factor $F^{(1)}_{\rm T2}(x_1,x_3,\mu)$ and $F^{(1)}_{\rm T3}(x_1,x_3,\mu)$
appeared in Eq.~(\ref{eq:ffnlop})
describe the NLO twist-2 and twist-3 contributions to the form factor $f^{+,0}(q^2)$ of the
$B_s \to K$ transition respectively,  and can be written in the following
form \cite{prd85-074004,cheng14a}:
\beq
F^{(1)}_{\rm T2}(x_1,x_3,\mu)&=& \frac{\alpha_s(\mu_f) C_F}{4 \pi}
\Biggl [\frac{21}{4} \ln{\frac{\mu^2}{m^2_{B_s}}}-(\frac{13}{2} + \ln{r_1}) \ln{\frac{\mu^2_f}{m^2_{B_s}}}
+\frac{7}{16} \ln^2{(x_1 x_3)}+ \frac{1}{8} \ln^2{x_1} \non
&&+ \frac{1}{4} \ln{x_1} \ln{x_3}
+ \left (- \frac{1}{4}+ 2 \ln{r_1} + \frac{7}{8} \ln{\eta} \right ) \ln{x_1}
+ \left (- \frac{3}{2} + \frac{7}{8} \ln{\eta} \right) \ln{x_3} \non
&& + \frac{15}{4} \ln{\eta} - \frac{7}{16} \ln^{2}{\eta}+ \frac{3}{2} \ln^2{r_1} - \ln{r_1}
+ \frac{101 \pi^2}{48} + \frac{219}{16} \Biggr ], \nonumber
\label{eq:ffnlot2}
\eeq
\beq
F^{(1)}_{\rm T3}(x_1,x_3,\mu)&=&\frac{\alpha_s(\mu_f) C_F}{4 \pi}
\Biggl [\frac{21}{4} \ln{\frac{\mu^2}{m^2_{B_s}}}
- \frac{1}{2}(6 + \ln{r_1}) \ln{\frac{\mu^2_f}{m^2_{B_s}}}
+ \frac{7}{16} \ln^2{x_1} - \frac{3}{8} \ln^2{x_3} \non
&& \hspace{-1cm}+ \frac{9}{8} \ln{x_1} \ln{x_3}
+ \left (- \frac{29}{8}+ \ln{r_1} + \frac{15}{8} \ln{\eta} \right ) \ln{x_1}
+ \left (- \frac{25}{16} + \ln{r_4} + \frac{9}{8} \ln{\eta} \right) \ln{x_3} \non
&& \hspace{-1cm}+ \frac{1}{2} \ln{r_1} - \frac{1}{4} \ln^{2}{r_4} + \ln{r_3}
- \frac{9}{8} \ln{\eta} - \frac{1}{8} \ln^{2}{\eta} + \frac{37 \pi^2}{32}+ \frac{91}{32} \Biggr ],
 \label{eq:ffnlot3}
\eeq
where $r_1=m^2_{B_s}/\xi_1^2$ with the choice of $\xi_1=25 m_{B_s}$, and we also set $r_4=1$.
For the $B \to \eta_c (P,V)$ decays,
the large recoil region corresponds to the energy fraction  $\eta  \sim \textit{O}(1-r^2_2)$.
The factorization scale $\mu_f$ is set to be the hard scales
\beq
t^{a}=\max(\sqrt{x_3 \eta } \, m_{B} ,1/b_1,1/b_3), \quad  {\rm or} \quad
t^{b}=\max(\sqrt{x_1 \eta } \, m_{B} ,1/b_1,1/b_3),
\eeq
corresponding to the largest energy scales in Fig.~\ref{fig:fig2}(a) and \ref{fig:fig2}(b), respectively.
 The relation between the scale $\mu$ and $\mu_f$ is defined as the same one in
Refs.~\cite{prd85-074004,fan2013,cheng14a}
\begin{eqnarray}
\mu = t_s (\mu_{\rm f})  = \left \{ \exp \left[ c_1 + \left(\ln
{m_{B}^2 \over \xi_1^2}  +{5 \over 4} \right)  \ln{\mu_{\rm f}^2
\over m_{B}^2 } \right ]  \, x_1^{c_2 } \, x_3^{c_3} \right \}^{2/21}
\, \mu_{\rm f}. \label{ts function}
\end{eqnarray}
The explicit expressions of the coefficients  $c_{1,2,3}$ in above equation can be
found in Refs.~\cite{prd85-074004,cheng14a}.

\section{Numerical results}\label{sec:n-d}

In the numerical calculations, the following input parameters will be used implicitly.
The masses, decay constants and QCD scales are in units of GeV \cite{pdg2018}
\beq
\Lambda_{\overline{\mathrm{MS}}}^{(f=5)} &=& 0.225, \; m_{B_{u,d}}= 5.280,\;  m_{B_{s}} =  5.37,
\; m_{\eta_c}=2.9834, \; m_c=1.275,\; m_b=4.8,\non
m_0^\pi & =& 1.4, \; m_0^K=1.9, \;   m_K=0.494, \; m_{\rho} = 0.77, \; m_{\omega}=0.78,\; m_{K^*}= 0.89,\; m_{\phi}=1.02, \non
 f_{\pi} &=& 0.13, \;  f_K = 0.16, \;  f_{\rho}=0.209,\; f_{\rho}^{T}=0.165, \;  f_{\omega}=0.195, \; f_{\omega}^{T}=0.145,\non
f_{K^*}&=&0.217,\; f_{K^*}^{T}=0.185,\; f_{\phi}=0.231, \; f_{\phi}^{T}=0.20, \;  f_B= 0.190,  \;  f_{B_{s}} = 0.23, \non
f_{\eta_c} &=& 0.42,\;  M_W = 80.42,\; \tau_{B^0} = 1.519 {\rm ps}, \;  \tau_{B^\pm} = 1.638 {\rm ps}, \; \tau_{B_s^0} = 1.510 {\rm ps}. \label{eq:para}
\eeq
For the CKM matrix elements, we adopt the Wolfenstein parametrization up to
${\cal O}(\lambda^5)$ with the updated parameters as presented in Ref.~\cite{pdg2018}
\beq
\lambda=0.22453, \; A=0.836\pm 0.015, \;\bar{\rho}=0.122^{+0.018}_{-0.017}, \;\bar{\eta}=0.355^{+0.012}_{-0.011}.
\label{eq:ckme}
\eeq

 For the considered two-body $B \to \eta_c (P,V)$ decays, the branching ratios can be expressed as:
\beq
\mathcal {B}(B \to \eta_c (P,V))=\frac{G_F^2\tau_{B}}{32\pi m_{B}}(1-r_2^2)|\cala  (B \to \eta_c (P,V))|^2,
\eeq
where $\tau_{B}$ is the lifetime of the $B_{u,d}$ or $B_s$ meson.

For  the two $B \to \eta_c \pi$ decay modes,  for example,  the LO and NLO PQCD predictions for their  CP-averaged branching ratios
(in units of $10^{-5}$ ) are the following:
 \beq
\calb(B^-  \to \eta_c \pi^-)&=& \left\{\begin{array}{ll} 0.88^{+0.43}_{-0.29}(\omega_B) \pm 0.12(a_i) ^{+0.53}_{-0.39}(t)^{+0.07}_{-0.09}({\rm CKM} ),  & {\rm LO},\\
1.64^{+0.52}_{-0.40}(\omega_B)^{+0.22}_{-0.23}(a_i) ^{+0.18}_{-0.14}(t)^{+0.09}_{-0.12}({\rm CKM}), & {\rm NLO}, \end{array} \right. \non
\calb(B^0  \to \eta_c \pi^0)&=& \left\{\begin{array}{ll} 0.43^{+0.20}_{-0.13}(\omega_B) \pm 0.06(a_i) ^{+0.25}_{-0.20}(t)^{+0.03}_{-0.04}({\rm CKM} ),  & {\rm LO},\\
0.76^{+0.24}_{-0.19}(\omega_B)^{+0.10}_{-0.11}(a_i) ^{+0.08}_{-0.06}(t)^{+0.04}_{-0.05}({\rm CKM}), & {\rm NLO}, \end{array} \right.
\label{eq:etacpi}
\eeq
where the first error is from the shape parameter $\omega_B=0.40\pm 0.04$ GeV;  the second error is from
the Gegenbauer moments such as $a_{2}^\pi=0.25\pm 0.15$ as given in Eq.~(\ref{eq:aim}).
The third error arises from the variation of the hard scale $t$ from $0.8 t$ to $1.2 t$, which characterize the
effects of the remaining higher order contributions.
The last error is the combined one from the uncertainties of the relevant CKM matrix elements as given in Eq.~(\ref{eq:ckme}).
Other remaining theoretical errors are  very small in size and can be neglected.
From  the numerical results in Eq.~(\ref{eq:etacpi}), one can see that  the scale dependence of the LO PQCD predictions is reduced significantly
after  the inclusion of the NLO corrections:
from $\sim 60\%$ of the central value at LO level to  $\sim 11\%$ at NLO for both decay channels.
For other considered decay modes, we also found the similar relations among the uncertainties from different sources.

In Table \ref{Tab:brexp}, we list the PQCD predictions for the CP-averaged branching ratios
of the considered sixteen $B \to \eta_c (P,V)$ decays together with currently available experimental
measurements for five decay modes \cite{lhcb17,Babar05,Babar07,Babar08,Babar04,belle03,belle18,pdg2018}.
The label ``LO"  denote the PQCD predictions at the full leading order,  while the label  ``+VC" means that the additional NLO vertex corrections are included.
The label ``NLO" means that the contributions from the NLO twist-2 and twist-3 corrections to the form factors of
$B  \to P$ transitions are also taken into account.
For $B  \rightarrow \eta_c V$ decays, unfortunately,
such NLO twist-2 and twist-3 corrections to $B \to V$ transition form factors are still not known.
In Table \ref{Tab:brexp}, we show the total theoretical uncertainties for the NLO PQCD predictions, obtained by adding the individual errors in quadrature.

As comparison,  we also listed the previous LO PQCD predictions for the four $B \to \eta_c (K, K^*)$ decays  as given in Ref.~\cite{pqcd05} and
the PQCD predictions for the two $B \to \eta_c K$ decays with the inclusion of the NLO vertex corrections as given in Ref.~\cite{cpc10}.
The central values of the theoretical predictions obtained from the QCDF approach \cite{QCDF03}, the Light-cone sum rule (LCSR) \cite{lcsr2}
 and  the final state interaction (FSI) ~\cite{fsi1} are also listed.
 Those currently available experimental measurements as given in PDG 2018 \cite{pdg2018} are also presented in last
column of Table  \ref{Tab:brexp}.

\begin{table}[tb]  
\caption{ The LO and NLO PQCD predictions for the CP-averaged branching ratios of  the
sixteen $B \to \eta_c (P,V)$ decays.  As a comparison, we also list the theoretical predictions as given in Refs.~\cite{pqcd05,cpc10,QCDF03,fsi1,lcsr2,npb924}
and the measured values as given in PDG 2018 \cite{pdg2018}.}
\label{Tab:brexp}
\begin{tabular*}{16.5cm}{@{\extracolsep{\fill}}l|lll|ll|l | l} \hline\hline
Modes & LO &  +VC  &  NLO &  PQCD$_{\rm LO}$  & PQCD$_{\rm NLO}$ & Others                                                & PDG \cite{pdg2018}\\ \hline
$ B^-    \to \eta_c K^- (10^{-4})                      $&2.56&    $5.58 $&$6.05^{+2.28}_{-1.65}   $& $2.34^{+2.43}_{-2.11}$\cite{pqcd05}&$5.9^{+2.5}_{-2.1}$\cite{cpc10}
 &$1.4-1.9$\cite{QCDF03}                        & $9.6\pm 1.1$     \\
$\bar B^0\to \eta_c \bar K^0 (10^{-4})     $&2.37&    $5.36 $&$5.61^{+2.12}_{-1.53}   $  &$2.19^{+2.13}_{-2.12}$\cite{pqcd05}&$5.5^{+2.3}_{-2.0}$\cite{cpc10}
&$1.4-1.9$\cite{QCDF03}  &$8.0\pm 1.2$     \\
$\bar B^0\to \eta_c \bar K^{*0} (10^{-4}) $&2.51&    $5.85$&$5.85^{+1.78}_{-1.32}   $&$2.64^{+2.71}_{-2.58}$\cite{pqcd05}&$-$
& $4.8-6.9$\cite{fsi1}  &$6.3\pm 0.9$     \\
$ B^-    \to \eta_c \bar K^{*-}  (10^{-4}) $      &2.71&    $6.31 $&$6.31^{+1.93}_{-1.42}  $&$2.82^{+2.91}_{-2.76}$\cite{pqcd05}& $-$
& $2.0\pm 0.1$\cite{lcsr2} &$10^{+5}_{-4}$     \\
$\bar B_s^0\to \eta_c \phi  (10^{-4})      $     &2.84&    $5.63 $&$5.63^{+1.86}_{-1.38}
$&$-$&$-$                  &$-$    & $ 5.0\pm 0.9$     \\ \hline \hline
$ B^-    \to \eta_c \pi^-  ( 10^{-5})            $&0.88&    $1.58 $&$1.64^{+0.60}_{-0.50} $      & $-$&$-$&$-$& $- $ \\
$\bar B^0\to \eta_c \pi^0 (10^{-5})          $&0.43&    $0.73 $&$0.76^{+0.27}_{-0.23}   $ & $-$&$-$&$-$& $- $ \\
$\bar B^0\to \eta_c \eta (10^{-5})  $&0.17&    $0.34 $&$0.38^{+0.16}_{-0.12}   $  & $-$&$-$&$-$& $- $\\
$\bar B^0\to \eta_c \eta^{\prime} (10^{-5})    $&0.12&    $0.23 $&$0.26^{+0.11}_{-0.08}   $ & $-$&$-$&$-$& $- $ \\
$ B^-    \to \eta_c \rho^- (10^{-5})            $&0.97&    $2.21 $&$2.21^{+0.64}_{-0.50}  $ & $0.85^{+0.46}_{-0.31}$\cite{npb924}&$-$&$-$& $- $ \\
$\bar B^0\to \eta_c \rho^0  (10^{-5})            $&0.45&    $1.01 $&$1.01^{+0.30}_{-0.23}   $   & $0.40^{+0.21}_{-0.14}$\cite{npb924}&$-$&$-$& $- $\\
$\bar B^0\to \eta_c \omega (10^{-5})            $&0.40&    $0.90 $&$0.90^{+0.27}_{-0.20}  $  & $-$&$-$&$-$& $- $\\ \hline \hline
$\bar B_s^0\to \eta_c \eta  (10^{-5})             $&5.4&    $15.1 $&$14.8^{+5.1}_{-4.4}   $   & $-$&$-$&$-$& $- $\\
$\bar B_s^0\to \eta_c \eta^{\prime} (10^{-5})    $&8.0&    $22.4 $&$22.1^{+8.1}_{-6.8}   $   & $-$&$-$&$-$& $- $\\
$\bar B_s^0\to \eta_c K^0 (10^{-5})               $&0.84&    $1.80 $&$1.94^{+0.78}_{-0.61}  $   & $-$&$-$&$-$& $- $ \\
$\bar B_s^0\to \eta_c K^{*0}  (10^{-5})         $&1.01&    $2.31 $&$2.31^{+0.77}_{-0.60}   $   & $-$&$-$&$-$& $- $\\ \hline\hline
\end{tabular*}
\end{table}

From the numerical results and the experimental data as listed in Table \ref{Tab:brexp},  we find the following points:
\begin{enumerate}
\item[(1)]
For all considered decays,  the NLO vertex corrections can provide large enhancements to the LO PQCD predictions of their branching ratios,
about $80\% -180\%$  in magnitude.
For the NLO Twist-2 and Twist-3 contributions, however,    play a minor role only:
resulting in an enhancement or a decrease less than$10\%$ to  $B \to \eta_c P$ decay  modes.
Among the five measured decays,  the central values of the LO PQCD predictions for their  decay rates are clearly much smaller than the measured ones.
The large NLO contributions can provide a great help for us to interpret the data as listed in last column of Table \ref{Tab:brexp}.
It is easy to see  that the NLO PQCD predictions for $\calb (B \to \eta_c (K,K^*,\phi) $  agree well  with the measured values  \cite{pdg2018}
within two standard deviations.   For  three $B \to \eta_c (K,K^{*-}) $ decays,  specifically,   the central values of the decay rates are smaller than the measured ones, and
there seems some space  left for still unknown higher order corrections or the non-perturbative contributions to these decays,
which would be further studied and tested in the future.

\item[(2)]
At the quark level, all  considered  decays can be classified into two types. The type-1 decays include the CKM-favored $B \to \eta_c (K, K^*)$ and
$B_s \to \eta_c (\eta_s,\phi)$ decays,    corresponding to the $b \to (c\bar{c})s$ transition at the quark level,
and  have the decay rates proportional to $|V_{cb}^*V_{cs} |^2\sim \lambda^4$.
The type-2 ones are  the  CKM-suppressed $B \to \eta_c (\pi,\eta_q,\rho,\omega)$ decays,  corresponding to the $b \to (c\bar{c})d$ transitions,
and have  the decay rates proportional to $|V_{cb}^*V_{cd} |^2\sim \lambda^6$.
The PQCD predictions for the branching ratios of the type-1 decays are  about $20-30$ times larger than the ones for type-2 decays mainly due to
the CKM enhancement  $|V_{cs}/V_{cd} |^2\sim \lambda^{-2}\approx  21$.

\item[(3)]
For the decays involving $\eta$ and $ \etar $ mesons,   the $\eta-\etar$ mixing can also help us to understand
the differences between the PQCD predictions for $\calb (B\to \eta_c (\eta,\etar) ) $ or  $\calb(B_s \to \eta_c (\eta,\etar) ) $.
Considering the $\eta-\etar$ mixing  as defined by Eq.~(\ref{eq:e-ep2}) , we have the expression for $\eta$ and $\etar$:
\beq
\eta &=&\eta_q \cdot  \cos\phi -  \eta_s \cdot \sin\phi, \non
\etar &=& \eta_q \cdot \sin\phi + \eta_s \cdot \cos\phi.
\label{eq:mix2}
\eeq
For the CKM-suppressed $\bar{B}^0 \to \eta_c (\eta,\etar)$ decays,   only the $d\bar{d}$ component of $\eta_q$ contributes, and we can define and
evaluate  the ratio $R_1$:
\beq
R_1&=&\frac{|A(\bar B^0 \to \eta_c \etar)|^2}{|A(\bar B^0 \to \eta_c \eta)|^2 }  \approx  \frac{\sin^2(\phi)}{\cos^2(\phi)} \approx 0.67.
\label{eq:ratio1}
\eeq
For the CKM-favered $\bar{B}_s^0 \to \eta_c (\eta,\etar)$ decays,   only the $\eta_s$ contributes, and we can define and
evaluate  the ratio $R_2$:
\beq
R_2&=&\frac{|A(\bar B_s^0 \to \eta_c \etar)|^2}{|A(\bar B_s^0 \to \eta_c \eta)|^2 }  \approx  \frac{\cos^2(\phi)}{\sin^2(\phi)} \approx 1/R_1 \approx 1.49.
\label{eq:ratio2}
\eeq
These two ratios could be measured and tested in the LHCb and  Belle-II experiments.
As a primary estimation for  the ratios $R_1$ and $R_2$,
the possible effects of the different phase space factors for $\eta$ and $\eta'$ meson  are not large in magnitude and have been neglected in this paper.

\end{enumerate}

Besides the decay rates, some ratios of the branching
fractions for  the decay modes involving $K$ and $K^*$ mesons have also been defined and measured
by the BaBar and Belle Collaborations \cite{Babar04,Babar08,belle03}.
As is well-known, one major advantage of studying the ratios of the branching ratios for the properly selected  pair of the decay modes
is the large cancelation of the theoretical uncertainties.
The three relative ratios measured by Babar and Belle \cite{Babar04,Babar08,belle03} and the corresponding PQCD predictions are the following:
\beq
R_3&=&\frac{{\cal B}( \bar{B}^0\to \eta_c \bar K^{0})}{ {\cal B}( \bar{B}^-\to \eta_c  K^-)}
= \left\{\begin{array}{ll}
0.93\pm 0.10,  & {\rm PQCD},\\ 0.87 \pm 0.15,  & {\rm BaBar},\\  \end{array} \right. \label{eq:ratio3}\\
R_4&=&\frac{{\cal B}( \bar{B}^0\to \eta_c \bar K^{*0})}{ {\cal B}( \bar{B}^0\to \eta_c \bar K^{0})}
= \left\{\begin{array}{ll}
1.04^{+0.07}_{-0.06},  & {\rm PQCD},\\  1.33^{+0.43}_{-0.49}, & {\rm Belle},\\ \end{array} \right.
\label{eq:ratio4}\\
R_5&=&\frac{{\cal B}( \bar{B}^0\to \eta_c \bar K^{*0} )}{ {\cal B}(
\bar{B}^-\to \eta_c K^-)}
= \left\{\begin{array}{ll} 0.96^{+0.08}_{-0.05},  & {\rm PQCD},\\ 0.62 \pm 0.08,  & {\rm BaBar}.\\ \end{array} \right.
\label{eq:ratio5}
\eeq
It is easy to see that the PQCD predictions for both $R_3$ and $R_4$ agree very well with the
measured values within one standard deviation. The theoretical errors of the PQCD predictions for the ratios $R_{3,4,5}$ are around ten percent,
which have been smaller than the uncertainties of currently available experimental measurements ( from $13\%$ to $37\%$)   \cite{Babar04,Babar08,belle03}.
The ratio $R_3$ is mainly governed by the difference between the lifetime of the $\bar B^0$ and $B^-$ mesons:
\begin{eqnarray}
R_3=\frac{{\cal B}( \bar{B}^0\to \eta_c \bar K^{0})}{ {\cal B}( \bar{B}^-\to \eta_c K^-)} \approx
\frac{\tau_{\bar B^0}}{\tau_{B^-}}\cdot \frac{|A(\bar{B}^0\to \eta_c \bar K^{0})|^2}{|A(B^-\to \eta_c K^-)|^2}
\approx \frac{\tau_{\bar B^0}}{\tau_{B^-}} \approx 0.93.
\end{eqnarray}
The ratio $R_4$ has  a dependence on the distribution amplitudes of the $K$ and $K^*$ mesons.
For the ratio $R_5$, the central value of our theoretical prediction is slightly larger than the measured one.
In fact, this ratio satisfy the relation of $R_5=R_3\cdot R_4$ by definition. These ratios will be tested by experiments
when more precise data from Belle-II and LHCb become available in the near future.

Analogous  to the ratio $R_1$,  we can also define the ratio $R_6$ for the decays involving $(\pi,\rho)$ mesons:
\begin{eqnarray}
R_6=\frac{{\cal B}(\bar B^0 \to \eta_c \pi^0)}{{\cal B}( B^- \to \eta_c \pi^-)}
   =\frac{{\cal B}(\bar B^0 \to \eta_c \rho^0)}{{\cal B}(B^- \to \eta_c \rho^-)}\approx \frac{1}{2}\cdot \frac{\tau_{\bar B^0}}{\tau_{B^-}}
   \approx 0.46.
\end{eqnarray}

Based on the similarity between $\bar B^0$ and $\bar B_s^0$ meson decays and the small SU(3) breaking effect,  it is reasonable for us to
define the ratio $R_7$ between $\calb(  \bar B_s^0\to \eta_c \bar K^{*0} )$ and  $\calb(  \bar B_s^0\to \eta_c \bar K^{0} )$ and
expect a similar PQCD prediction with $R_4$. Direct numerical calculation tell us that:
\begin{eqnarray}
R_7= \frac{{\cal B}( \bar B_s^0\to \eta_c \bar K^{*0})}{ {\cal B}( \bar B_s^0\to \eta_c \bar K^{0})} \approx 1.19,
\end{eqnarray}
which is actually close to $R_4=1.04^{+0.07}_{-0.06}$.

\begin{table}[t]  
\caption{ The PQCD predictions for the direct CP violation $\cala^{\rm dir}_{\rm CP}$ $ (\%)$ and mixing induced CP violation
$\cala^{\rm mix}_{\rm CP} $ ($\%$) of the considered $B \to \eta_c (P,V)$ decays.
The only currently available data is from BaBar \cite{Aubert:2009aw,pdg2018}
as listed in Eq.~(\ref{eq:acpe1}). }
\label{Tab:cpv1}
\begin{tabular*}{7cm}{@{\extracolsep{\fill}}l|l|l} \hline\hline
{\rm Modes}&{\rm $A^{\rm dir}_{CP}$} &{\rm $A^{\rm mix}_{CP} $}\\ \hline
$ B^-    \to \eta_c  K^{*-}                  $&$ 0.06\pm 0.02$&$-$\\
$ B^-    \to \eta_c  K^-                     $&${0.09\pm 0.03}$&$-$\\
$ B^-    \to \eta_c \pi^-                  $&$1.6^{+0.7}_{-0.5}$&$-$\\
$ B^-    \to \eta_c \rho^-                 $&$-1.7^{+0.3}_{-0.2}$&$-$\\  \hline
$\bar B^0\to \eta_c  K_S               $&$0.09\pm 0.03$&                                         $71\pm 1$\\
$\bar B^0\to \eta_c \pi^0                  $&$1.6^{+0.7}_{-0.5}$&$-73\pm 2 $\\
$\bar B^0\to \eta_c \eta                              $&$1.8\pm 0.7 $&$-76\pm 2   $\\
$\bar B^0\to \eta_c \eta^{\prime}          $&$1.8\pm 0.7$&$-76\pm 2   $\\ \hline
$\bar B^0\to \eta_c \bar K^{*0}              $&$0.06\pm 0.02$&$-$\\
$\bar B^0\to \eta_c \rho^0                 $&$-1.7^{+0.3}_{-0.2}$&$-73^{+3}_{-2}$\\
$\bar B^0\to \eta_c \omega                 $&$-1.7^{+0.3}_{-0.2}$&$-73^{+3}_{-2}$\\
\hline\hline
\end{tabular*}
\end{table}

\begin{table}[t]  
\caption{ The PQCD predictions for the CP violating asymmetries ($\%$)
of the considered $\bar{B}^0_s \to \eta_c (P,V)$ decays. }
\label{Tab:cpv2}
\begin{tabular*}{8cm}{@{\extracolsep{\fill}}l|l|l|l} \hline\hline
{\rm Modes}&{\rm $A^{\rm dir}_{CP}$} & {{\rm $A^{\rm mix}_{CP}$}} & $ H_f$ \\ \hline
$\bar B_s^0\to \eta_c \eta                 $&${-0.08\pm 0.03}$&                                         $-3.8\pm 0.1$ &$\sim 99$\\
$\bar B_s^0\to \eta_c \eta^{\prime}        $&${-0.08\pm 0.02}$&                                         $-3.8\pm 0.1$ &$\sim 99$\\
$\bar B_s^0\to \eta_c \phi                 $&$0.05^{+0.02}_{-0.01}$&                                         $-3.8\pm 0.1$ & $\sim 99$\\ \hline
$\bar B_s^0\to \eta_c K^{*0}               $&$-1.4^{+0.3}_{-0.2}$&$-$ & $\sim 99$ \\
$\bar B_s^0\to \eta_c K_S                  $&$1.6^{+0.7}_{-0.6}$& $6\pm 2 $ &$\sim 99$\\ \hline\hline
\end{tabular*}
\end{table}

Now we turn to the evaluations of the CP-violating asymmetries for the considered decay modes.
For the charged $B^\pm$ meson decays, there exists the direct CP violation asymmetry $\cala ^{\rm dir}_{CP}$ only, which can be
defined as usual:
\begin{eqnarray}
A^{\rm dir}_{CP}=\frac{|A(B^- \to f)|^2-|A( B^+ \to \bar{f})|^2}{|A(B^- \to f)|^2+|A( B^+ \to \bar{f})|^2}.
\label{eq:acpd1}
\end{eqnarray}

For the neutral $B^0$ decays, the  mixing effects should be taken into account. For $B^0$ decays,  the very small ratio
$\Delta \Gamma_d/\Gamma_d =-0.002\pm 0.010 $ \cite{hfag2016} can be neglected safely.
The direct and mixing-induced  CP violation $\cala^{\rm dir}_{\rm CP}$ and
$\cala^{\rm mix}_{\rm CP}$ can then be defined in the following form:
\beq
\cala^{\rm dir}_{CP}=\frac{|\lambda_f|^2-1}{1+|\lambda_f|^2}, \quad
\cala^{\rm mix}_{CP}=\frac{2 {\rm Im} (\lambda_f)}{1+ |\lambda_f|^2},
\label{eq:acpd2}
\eeq
with the  CP violating parameter $\lambda_f$:
\beq
\lambda_f=\eta_f e^{-2i\beta} \frac{\langle f|H_{eff}| \bar{B}^0 \rangle}{\langle f |H_{eff}| B^0\rangle },
\label{eq:lf1}
\eeq
where $\eta_f=\pm 1$  for a CP-even or CP-odd final state $f$,  and  $\beta=\arg\left [-(V_{cd}V_{cb}^*)/(V_{td}V_{tb}^*)  \right]$ is the phase
angle for $B^0$ system.

For the neutral $B_s^0$ decays, the  ratio $\Delta \Gamma_s/\Gamma_s \approx 0.13 $ \cite{hfag2016}
is large and should  be
taken into account in our calculations for the CP violating asymmetries.
For $B_s^0$ decays, the CP asymmetries $\cala^{\rm dir}_{\rm CP}$,
 $A^{\rm mix}_{CP}$ and $H_f$ are constrained physically by the relation
$|\cala^{\rm dir}_{\rm CP}|^2 + |A^{\rm mix}_{CP}|^2 +|H_f|^2=1$, and can be defined
in the usual way:
\beq
\cala^{\rm dir}_{\rm CP}=\frac{|\lambda_f|^2-1}{1+|\lambda_f|^2}, \quad
\cala^{\rm mix}_{\rm CP}=\frac{2 {\rm Im} (\lambda_f)}{1+ |\lambda_f|^2},  \quad  H_f=\frac{2 {\rm Re} (\lambda_f)}{1+ |\lambda_f|^2},
\eeq
with the  CP violating parameter $\lambda_f$:
\beq
\lambda_f=\eta_f e^{-2i\beta_s} \frac{\langle f|H_{eff}| \bar{B}_s^0 \rangle}{\langle f |H_{eff}| B^0_s \rangle},
\label{eq:lf2}
\eeq
here $\beta_s=\arg\left[-(V_{ts}V_{tb}^*)/(V_{cs}V_{cb}^*)  \right] $
is the phase angle for $B_s^0$ system.

Among the sixteen $B \to \eta_c (P,V)$ decays considered in this work,
only the CP asymmetries of the decay $B^0 \to \eta_c  K_S^0$ have been measured now \cite{Aubert:2009aw,pdg2018}, as listed in Eq.~(\ref{eq:acpe1}).
In  Table~\ref{Tab:cpv1} and \ref{Tab:cpv2}, we list the PQCD predictions for the CP violating asymmetries of the considered $B_{u,d}$ and $B_s^0$
decay modes respectively.
The errors here are defined in the same way as those for the branching ratios. For the direct CP asymmetries,
the error from the wave function parameters is largely cancelled between the numerator and denominator.
For the mixing induced CP asymmetries, the errors from the input hadronic quantities and CKM matrix elements are actually very small,
and we only list the total errors by adding the individual errors in quadrature.
From the numerical results as listed  in Table~\ref{Tab:cpv1} and \ref{Tab:cpv2},  one can see the following points:
\begin{itemize}
\item[(1)]
 For the seven $b \to c\bar{c}s$ transition decays of $B \to \eta_c (K^-,K^{*-}, K^{*0}, K_S)$ and $\bar{B}^0_s \to \eta_c (\eta^{(\prime)},\phi)$,
their direct CP asymmetries will be zero when the CKM matrix at the leading order  are used in the calculations, due to the absence of the weak phase
in the relevant CKM factor $V_{cb}V_{cs}^*$ and $V_{tb}V_{ts}^*$ in their decay amplitudes as given in Eq.~(\ref{eq:aa10})
\footnote{  In the Wolfenstein parametrization up to LO $({\cal O}(\lambda^3))$,    we have the real $V_{cb}=A \lambda^2$,
$V_{cs}=1-\frac{\lambda^2}{2}$, $V_{tb}=1$ and $V_{ts}=-\frac{\lambda^2}{2}$  \cite{pdg2018}.   }.
When we use the NLO CKM matrix elements
\footnote{ In the Wolfenstein parametrization up to NLO$({\cal O}(\lambda^5))$, we have $V_{ts}=-A \lambda^2+A \lambda^4(\frac{1}{2}-\rho-i\eta)$,
while other three CKM elements $V_{cs}=1-\frac{\lambda^2}{2}-\lambda^4(\frac{1}{8}+\frac{A^2}{2})$, $V_{cb}=A\lambda^2$ and
$V_{tb}=1-\frac{A^2\lambda^4}{2}$ are still pure real number\cite{pdg2018}.},
however,  a small imaginary part appears in $V_{ts}$  and  the resultant direct CP asymmetries become nonzero but  tiny in size
due to the strong suppression factor ( $ \propto \lambda^4$).
For $B^-\to \eta_c  K^{*-}$ decay, for example, we found a PQCD prediction for direct CP violation at the $10^{-4}$ level:
$\cala^{\rm dir}_{\rm CP}(B^-\to \eta_c  K^{*-})=6\times 10^{-4}$,  as presented in  Table~\ref{Tab:cpv1}.
For the  remaining nine $b \to c\bar{c} d$ transition decays,
the corresponding weak phase is also rather small in size,  their direct CP violating parameters are consequently less than $ 2\%$ in magnitude
as listed in Table~\ref{Tab:cpv1} and \ref{Tab:cpv2}.

\item[(2)]
For the neutral $B^0/B_s^0$ meson decays,  because of the very small  $\cala^{\rm dir}_{CP}$,
the  mixing induced CP asymmetries $\cala^{\rm mix}_{CP}$  are approximately proportional to the $\sin2\beta$ or  $\sin2\beta_s$,
specially for the decays of $\bar B^0\to \eta_c \bar K^0$ and $\bar B_s^0\to \eta_c (\eta^{(\prime)},\phi)$.
The PQCD predictions  agree very well  numerically with the current world average values $\sin2\beta$ and $-2\beta_{s}$ \cite{pdg2018}.

\item[(3)]
For $B^0 \to \eta_c  K_S^0$ decay,   the PQCD  predictions for both the direct and mixing induced CP asymmetries as listed in Table~\ref{Tab:cpv1}  do agree
well with  the measured values  as given in Eq.~(\ref{eq:acpe1})  within errors.
It is easy to see that the direct CP violation has not been seen by experiment up to now.
Since the PQCD prediction $\cala^{\rm dir}_{ CP}( B^0 \to \eta_c K_S^0)\approx 0.1\%$ is very small
in size, any observation of the large direct CP asymmetries for this decay mode
will be a possible signal for the new physics beyond  SM.
Besides  the  measured  large value $\cala^{\rm mix}_{ CP}( B^0 \to \eta_c K_S^0)=0.93\pm 0.17$ ,
fortunately,  the large mixing induced CP asymmetry $ (\sim 70\%)$ for
other decays with similar $b \to c\bar{c}d$ transition are also measurable
in the near future LHCb and Belle-II experiment.
\end{itemize}


\section{Summary}\label{sec:4}

In summary, we studied the  sixteen $B \to \eta_c (P,V)$ decays by employing the PQCD factorization approach with the inclusion
of the all currently known NLO contributions. We calculated the branching ratios and CP-violating asymmetries of the considered
decay modes, defined several ratios of the decay rates, and compared our PQCD predictions with the measured values or the previous theoretical predictions
based on the PQCD approach or other methods.

From our numerical calculations and phenomenological analysis, we found the following points:
\begin{itemize}
\item[(1)]
The NLO vertex corrections can provide about $80\% -180\%$  enhancements to the LO PQCD predictions of the branching ratios of the  considered
decay modes. The NLO Twist-2 and Twist-3 contributions to the form factors of $B \to P$ transitions,
however, can only provide  a relatively small change, less than $10\%$ to  $B \to \eta_c P$ decay modes.
For the five measured decays, the NLO PQCD predictions for  their decay rates
are well consistent with  currently available
experimental measurements within $2\sigma$ errors, and
the NLO contributions do play an important role in understanding the data,
as can be seen easily from the numerical results in Table \ref{Tab:brexp}.

\item[(2)]
We defined seven ratios of the branching ratios for  properly selected pairs of  considered decay modes.
For the three measured ratios $R_{3,4,5}$,  the PQCD predictions  agree well with currently available
BaBar and Belle measurements. For other four ratios $R_{1,2}$ and $R_{6,7}$, the PQCD predictions also agree
with the general expectations and will be tested by the future experiments..

\item[(3)]
For all considered decays, the PQCD predictions for the CP-violating asymmetries agree with the general expectations.
For the only measured $B^0 \to \eta_c  K_S^0$ decay,   the PQCD  predictions for both the direct and mixing induced CP asymmetries
do agree very well with  the measured values within errors:
\beq
\cala^{\rm dir}_{CP}( B^0 \to \eta_c K_S^0)&=& \left\{\begin{array}{ll}
(9\pm 2)\times 10^{-4},  & {\rm PQCD},\\  0.08 \pm 0.13,  & {\rm PDG2018},\\  \end{array} \right. \label{eq:acpks1} \\
\cala^{\rm mix}_{CP}( B^0 \to \eta_c K_S^0)&=& \left\{\begin{array}{ll}
0.71\pm 0.01,  & {\rm PQCD},\\
0.93 \pm 0.17,  & {\rm PDG2018}.\\  \end{array} \right. \label{eq:acpks2}
\eeq
The large mixing induced CP asymmetries $ (\sim 70\%)$ for other similar CKM-suppressed $b \to c\bar{c}d$
transition decays could be measured  in the future LHCb and Belle-II experiments.
\end{itemize}

\begin{acknowledgments}

This work is supported by the National Natural Science
Foundation of China under Grants  No.~11775117,  11875033  and No.~11765012,
by the Qing Lan Project of Jiangsu Province (Grant No.~9212218405),
and by the Research Fund of Jiangsu Normal University(Grant No.~HB2016004).

\end{acknowledgments}



\end{document}